\documentclass[%
 aip,
 amsmath,amssymb,
 reprint,%
]{revtex4-1}
\usepackage{amsfonts}

\usepackage{bm}%
\usepackage[]{hyperref}
\usepackage{graphicx}
\usepackage{amsmath}
\usepackage{siunitx}
\usepackage{soul}
\usepackage{amssymb}
\makeatletter
\usepackage{rotating}
\usepackage{url}
\usepackage[normalem]{ulem}
\usepackage{anyfontsize}
\usepackage{blindtext}
\usepackage[utf8]{inputenc}
\usepackage{xr}
\usepackage{lipsum}
\usepackage{ccaption}

\usepackage{floatrow}
\usepackage[font= large,label font=bf,labelformat=simple]{subfig}

\usepackage[T1]{fontenc}
\usepackage{lipsum}
\usepackage[table]{xcolor} 
\usepackage{tabularx}
\begin{document}

\title{Self-generated persistent random forces drive phase separation in growing tumors}
\author{Sumit Sinha}\affiliation{Department of Physics, University of Texas at Austin, TX 78712}
\author{D. Thirumalai}\affiliation{Department of Chemistry, University of Texas at Austin, TX 78712}

\begin{abstract}
A single solid tumor, composed of nearly identical cells,  exhibits heterogeneous dynamics. Cells dynamics in the core is glass-like whereas those in the periphery undergo diffusive or super-diffusive behavior. Quantification of  heterogeneity using the mean square displacement or the self-intermediate scattering function, which involves averaging over the cell population, hides the complexity of the collective movement. Using the t-distributed stochastic neighbor embedding (t-SNE), a popular unsupervised machine learning dimensionality reduction technique,  we show that  the phase space structure of an evolving colony of cells, driven by cell division and apoptosis, partitions into nearly disjoint sets composed principally of core and periphery cells. The non-equilibrium phase separation is driven by the differences in the persistence of self-generated active forces induced by cell division.  Extensive heterogeneity revealed by t-SNE paves way towards understanding the origins of intratumor heterogeneity using experimental imaging data. 
\end{abstract}

\pacs{}

\maketitle
Intratumor heterogeneity (ITH), a pervasive phenomena across cancers, is a major hurdle in developing effective treatment \cite{Heppner83CancerMetastasisRev,mcgranahan2015biological}. ITH refers to the coexistence of genetically or phenotypically distinct cells within a single tumor \cite{hinohara2019intratumoral}. A source of ITH is genetic variations. Indeed, multi-region sequencing have revealed widespread genetic diversity within tumors \cite{gerlinger2012intratumor, gerlinger2014, harbst2016, debruin2014, zhang2014intratumor, cao2015}. Stochastic variations due to differences in cancer microenvironment, which results in vastly different dynamics of cells in distinct regions of an evolving solid tumor, could also give rise to ITH.  Evidence for the dynamically-driven ITH have emerged recently from several imaging studies, which have mapped out the phenotypic properties (such as shape and size) in three dimensional tumors spheroids \cite{richards20184d,valencia2015collective, kang2020tumor, han2019cell}. The growth of tumor spheroids are monitored by embedding them in a collagen matrix \cite{richards20184d,valencia2015collective, kang2020tumor, han2019cell,palamidessi2019unjamming}.  Direct imaging reveals that the dynamics of  cells in the tumor core differs dramatically compared to cells in the periphery \cite{valencia2015collective, kang2020tumor, sinha2020spatially, han2019cell}, a clear signature of dynamical ITH, which we abbreviate as DITH. A characteristic of DITH is that the material properties  of the cells are unaltered, implying that heterogeneity arises solely from microenvironment fluctuations. In this sense, DITH is reminiscent of dynamic heterogeneity in supercooled liquids that undergo glass transition~\cite{kirkpatrick2015colloquium,Berthier20JCP}.  

Previously we showed that the cell dynamics in a growing multicellular spheroid (MCS) is spatially heterogeneous \cite{malmi2018cell, sinha2020spatially}, which implies that cells in the core (periphery) exhibit sub-diffusive (super-diffusive) motion. These characteristics were first observed in imaging experiments tracking the displacement of cells moving in a collagen matrix, and recently in other studies as well \cite{kang2020tumor, han2019cell}. However, characterizing the dynamics using conventional ensemble average measures, such as mean squared displacement or the self-intermediate scattering function, hide the rich dynamics, the cause of DITH. 

Can we infer DITH directly from the cell trajectories in an evolving tumor?  Computer simulations of physical models for evolving cells, and more importantly, direct imaging can be used to generate the needed trajectories. Here, we show that the  t-distributed stochastic neighbor embedding (t-SNE), a popular unsupervised machine learning technique for analyzing big data, is ideally suited to answer the question posed above. The t-SNE method is among the best dimensionality reduction technique \cite{maaten2008visualizing,van2009learning,van2012visualizing,van2014accelerating}, allowing us to visualize the emergent heterogeneous dynamics {\it without any} inherent bias in the trajectory analysis. It has  been extensively used in various areas ranging from genomics \cite{kobak2019art, linderman2019fast}, neuroscience \cite{dimitriadis2018t} to condensed matter physics \cite{carrasquilla2017machine, ch2018unsupervised,day2019glassy}. 

We performed t-SNE on data generated using simulations of an  expanding tumor spheroid model \cite{malmi2018cell,malmi2019dual,sinha2020spatially,samanta2020far}. The results revealed massive dynamical heterogeneity that depends on the radial distance from the tumor center, which accords well with the conclusions in recent experiments \cite{valencia2015collective, kang2020tumor, han2019cell}. t-SNE also resolves the dynamical phase space structure of cells in the core and periphery. Division of the dynamical phase space structure primarily into two disjoint sets is a consequence of differences in the persistence of self generated active force (SGAF), which our model is dynamically generated due to an imbalance in the cell division and apoptosis rates. The cells in the periphery experience highly persistent forces that is predominantly radially outward. 

\textit{t-SNE:}
For completeness, we provide a brief description of the t-SNE method \cite{maaten2008visualizing,van2009learning,van2012visualizing,van2014accelerating}. Let us consider a $n$ dimensional vector, $\{{\bf x}_1, {\bf x}_2.....{\bf x}_n \}$, which in our case are the time dependent cell positions or forces experienced by the cells. The components of ${\bf x}_i$s are $\{x_{i,1}, x_{i,2} ..... x_{i,D}\}$ with $D \gg 1$). The t-SNE projects ${\bf x}_i$s onto a low dimensional (usually 2 or 3) space ${\bf y}_i$s while being faithful to the information content in the  high dimensional space.  

To determine the ${\bf y}$s, a joint probability $p_{ij}$ in high dimensional space, measuring the likelihood that points ${\bf x}_i$ and ${\bf x}_j$ are close to each other, is constructed. Following the standard practice, we take $p_{ij}=\frac{p_{i|j}+p_{j|i}}{2n}$ where the conditional probabilities $p_{j|i} \propto exp({-\frac{||{\bf x}_i-{\bf x}_j||^2}{2\sigma_i^2}})$, where ||....|| is a measure of distance, $\sum_{i,j}p_{ij}=1$, and $p_{i|i}$ is set to zero. The variance $\sigma_i^2$ is chosen such that the perplexity ($\mathcal{P}_i$) of the distribution is given by,
\begin{equation}
   \mathcal{P}_i=2^{-\sum_{j\neq i} p_{j|i}log_2p_{j|i}}.
\end{equation}
The perplexity  is independent of $i$ ($\mathcal{P}_i = \mathcal{P}$). The maximum perplexity can be $(n-1)$ which resulting in $\sigma_i=\infty$, which would lead to a uniform distribution (i.e $P_{j|i}=\frac{1}{n-1}$). The perplexity value, which can be interpreted as the number of effective neighbors, influences the outcome of t-SNE \cite{wattenberg2016use}.

The joint probability $q_{ij}$, measuring the likelihood that points ${\bf y}_i$ and ${\bf y}_j$ are in proximity, is  t-distributed (i.e $q_{ij}\propto [1+||{\bf y}_i-{\bf y}_j||^2]^{-1}$) with $q_{ii}=0$ and $\sum_{i,j}q_{ij}=1$. To compute the ${\bf y}$s, we use the Kullback-Leibler divergence ($\mathcal{L}=\sum_i\sum_j p_{ij}log \frac{p_{ij}}{q_{ij}}$) as a loss function ($\mathcal{L}$) to minimize the difference between $p_{ij}$ and $q_{ij}$. We determine ${\bf y}_i$s by minimizing  $\mathcal{L}$ using a gradient method. The gradient minimization is numerically implemented using the updating scheme, 
\begin{equation}
    {\bf y}_i[t]= {\bf y}_i[t-1]+\eta\frac{\partial \mathcal{L}}{\partial {\bf y}_i} + \alpha(t)\bigg({\bf y}_i[t-1]-{\bf y}_i[t-2]\bigg).
\label{eqmotion}    
\end{equation}
In eq.\ref{eqmotion} $\eta$ is the learning rate, and $\alpha (t)$ is the momentum term that is included to speed up the optimization. The ${\bf y}_i[0]$s are sampled from a normal distribution of mean 0 and variance 0.0001.

The parameters in the the t-SNE algorithm are $\mathcal{P}$, $\eta$, the momentum ($\alpha$), and number of iterations. We used $\eta$=200,  $\alpha(t)=0.5$ for $t\leq 250$ and $\alpha=0.8$ for $t\geq 250$. We performed 2,000 iterations. The perplexity is varied depending on the situation.  We projected three  large data sets onto two dimensions with coordinates tSNE1 and tSNE2. 


{\it Position data:} We collected the time traces of $\approx 5,000$ cells (x, y and z coordinates) between time interval $T_{w1}=\tau \leq t \leq 11\tau$, where $\tau=15~hrs$ represents the cell cycle time (see \cite{malmi2018cell,sinha2020spatially} for details). The cell positions were recorded every $500~s$. The sampling rate was chosen to roughly mimic the frame rate (one per $14~mins$) of microscopy measurements in experiments \cite{valencia2015collective,han2019cell}. The trajectory obtained from simulations were divided into 1,080 ($\frac{T_{w1}}{500s}$) time windows. Each time window can be thought of as a dimension. Therefore, the trajectories of each cell resides in 1,080 dimensions. In each time window $t_i$, a cell is displaced by $|\delta x (t_i)|=|x(t_{i+1})-x(t_i)|$ (similarly for y and z coordinates). Here, $|...|$ represents the absolute sign. Thus, for each cell we have $1080~(t_i,|\delta x_i|)$ pairs. Before applying the t-SNE, for each cell, $1080 |\delta x_i|$s were sorted from the smallest to the largest value. 

{\it Force data:} We also used time traces of forces on individual cells ($\approx 5,000$ cells) in the t-SNE analysis.  Forces, with $F_x, F_y$ and $F_z$, were recorded every $\approx 10~mins$ between time interval $T_{w1}=\tau\leq t \leq 11 \tau$.  

{\it Interpenetration data:} This data set contains the interpenetration  distances for $\approx 5,000$ cells that were present in the simulation for time interval $T_{w1}$. The interpenetration of the $i^{th}$ cell at time $t_k$ is given by, $h_i(t_k)=\frac{1}{NN(i,t_k)}\sum_{j=1}^{NN(i,t_k)} h_{ij}(t_k)$, where $h_{ij}(t_k)=max\{0, R_i(t_k)+R_j(t_k)-|{\bf r}_i(t_k)-{\bf r}_j(t_k)|\}$. Here, $R_j(t_k)$ (${\bf r}_j(t_k)$) is the radius (position) of the $j^{th}$ cell at time $t_k$. $NN(i,t_k)$ is the number of nearest neighbors of the $i^{th}$ cell at time $t_k$. For each cell we have $1080~(t_k,h_k)$ pairs.

\floatsetup[figure]{style=plain,subcapbesideposition=top}
\begin{figure}
\includegraphics[width=1\linewidth] {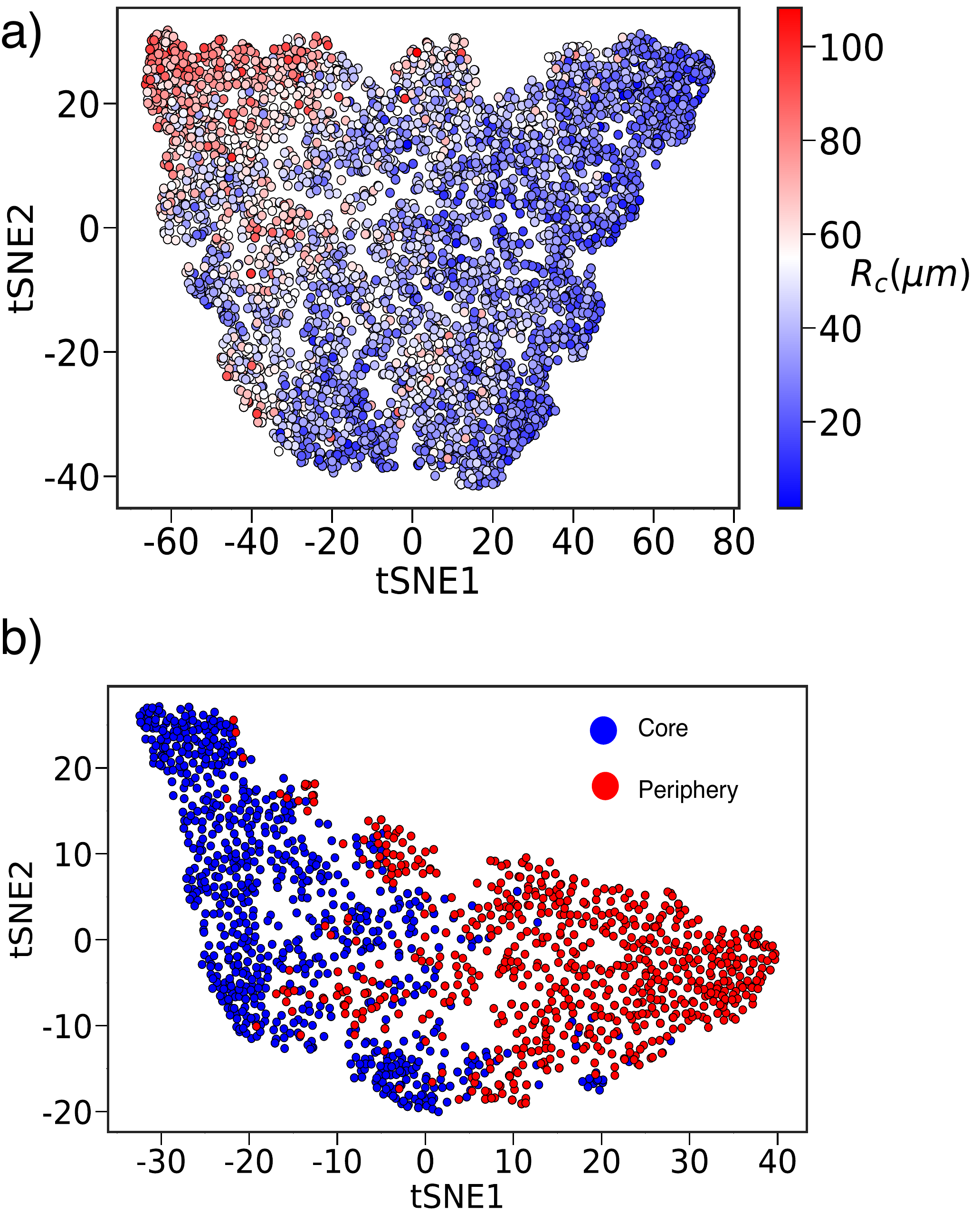}\label{fluid_contd}
\caption{{\bf Heterogeneity is radially dependent in an evolving tumor spheroid}. {\bf (a)} Cell trajectories ($\approx 5,000$ ) projected onto tSNE1 and tSNE2. Each dot represents a cell. The cell label depends on the radial distance from the center of the tumor ($R_c$). The color gradient is suggestive of extensive dynamical heterogeneity. The blue color represents cells closer to the center whereas cells colored in red are farther away from the center (see the $R_c$ scale on the right). {\bf (b)} Cell trajectories sampled from the core ( blue dots) and periphery ( red dots). There are 1580 (946) core (periphery) cells. Projection of cell trajectories onto t-SNE coordinates  shows that the cells in the two regions have resolvable dynamical phase space structure, by which we mean that the data cluster according the motilities of the cells. This is easily seen from the separation of the red and the blue dots.  }
\label{contd_fluid}
\end{figure}

\textit{Results:} 
Figure \ref{contd_fluid}a shows the clustering obtained when the trajectories of cells ($1080~(t_i,|\delta x_i|)$) are projected onto tSNE1 and tSNE2 ($\mathcal{P}=100$). In figure \ref{contd_fluid}, each dot represents a single cell and are colored depending on their distance from the center of the tumor, $R_c$. It should be emphasized that in performing t-SNE we did not use the information of the cell distance from the tumor center. The colors aid to visualize the cells. The results in figure \ref{contd_fluid}a show that there is pattern in the way the dots are arranged. Majority of the red dots (cells farthest from tumor core) are at one end, with blue dots at the other end (cells closer to the core). In the other words, there is a dynamic phase separation, which we show below is consequence of cell division and apoptosis. The partitioning into two disjoint patterns in figure \ref{contd_fluid}a implies that the dynamics of the cells is dependent on their distance from the center of tumor as noted in experiments \cite{valencia2015collective, kang2020tumor}. However, the boundary between the two regions (roughly high and low density) is not sharp.  The t-SNE method, based on machine learning, is able to delineate massive heterogeneity in a single tumor with identical cells, which is hidden in observables like mean squared displacement or self-intermediate scattering function \cite{malmi2018cell,sinha2020spatially}.  Thus, unbiased analyses of the cell trajectories are required to shed light on the origin of DITH in solid tumors \cite{kirkpatrick2015colloquium,Alemendro13ARPathol, li2019share}.

In a recent experiment \cite{kang2020tumor}, the solid tumor was divided into two core and periphery regions. It was shown that the tumor core (periphery) exhibits sub-diffusive (super-diffusive) dynamics, implying that  the cells explore distinct non-overlapping regions of phase space dynamically. In order to assess if  t-SNE separates  the dynamics in the two regions, we collected position data of core ($R_c <30~\mu m$) and periphery ($R_c >60~\mu m$). We applied the SNE algorithm on this mixed data set.   The  figure \ref{contd_fluid}b shows the t-SNE clustering of cells belonging to core (blue dots) and periphery (red dots) ($\mathcal{P}=100$). It is clear from figure \ref{contd_fluid}b that the red and blue cells are approximately phase separated. The distinct dynamical phase space explored by the cells in the core and periphery, as illustrated by figure \ref{contd_fluid}b, sheds light on the non-equilibrium phase separation between tumor core and periphery \cite{kang2020tumor}.

\floatsetup[figure]{style=plain,subcapbesideposition=top}
\begin{figure}
\includegraphics[width=1\linewidth] {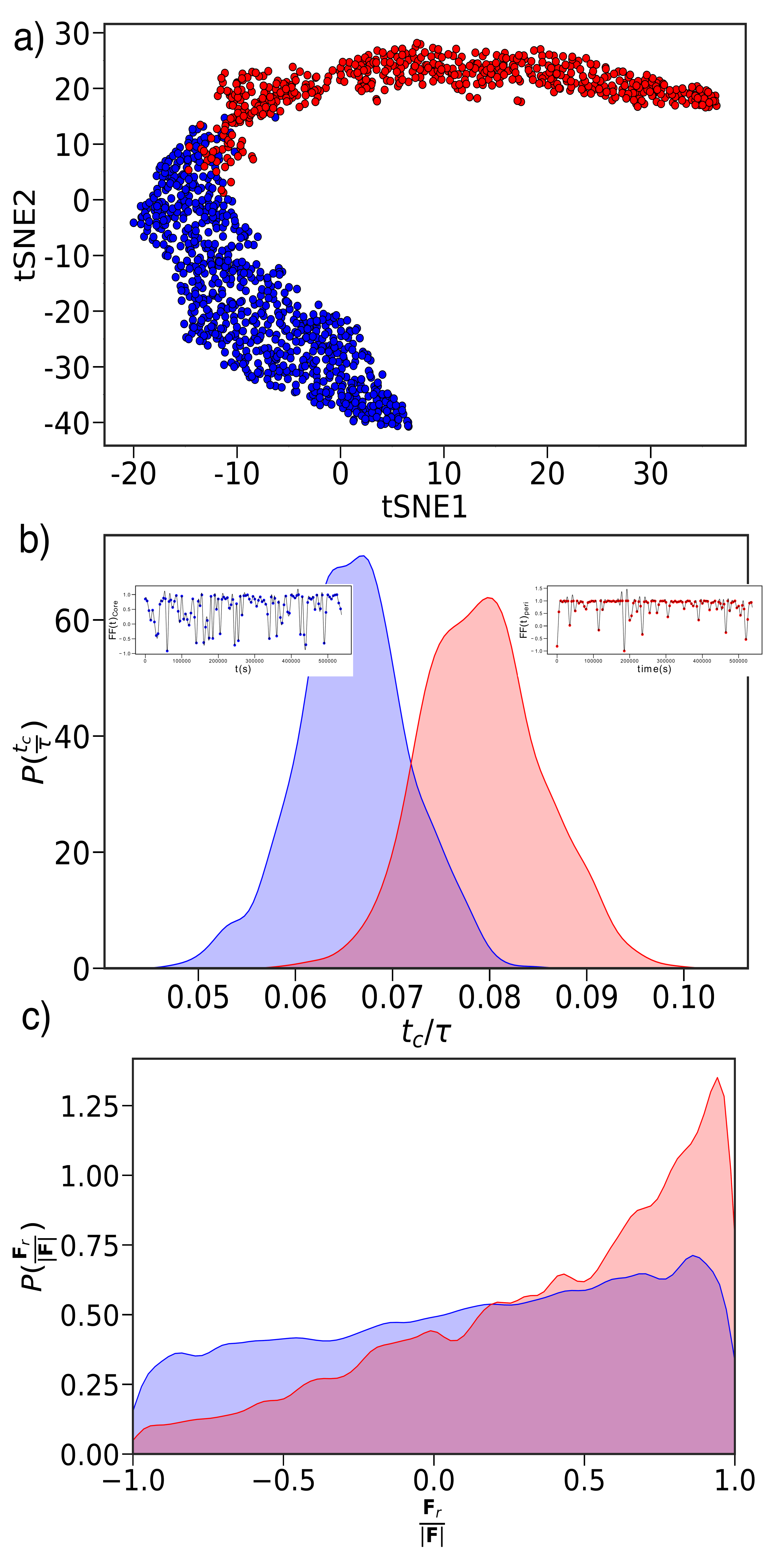}
\caption{{\bf Self generated force due to cell division and apoptosis}. {\bf (a)} t-SNE projection of $FF(t)$ for cells in the core (blue) and periphery (red). The cells in the two regions cluster into distinct regions. {\bf (b)} Distribution of $\frac{t_c}{\tau}$ obtained using Gaussian Process Regression for cells in the core (blue) and periphery (red). Cells in the periphery have higher  persistence times (mean value is $\frac{t_c}{\tau}=0.08$) compared to cells in the core (average is $\frac{t_c}{\tau}=0.06$). Inset on the left (right) shows the FF(t) fit using GPR for cells in core (periphery).  The blue (red) dots correspond to cells in the core (periphery) and the solid black line is the GPR fit. {\bf (c)} Probability distribution of $\frac{{\bf F}_r}{|{\bf F}|}$ for cells in the core (blue) and periphery (red). Cells in the periphery experience force predominantly in the radial direction.}
\label{force_cell}
\end{figure}

Phase separation of cells  into core and periphery is a consequence of self-generated active forces (SGAFs) arising from cell division. The SGAF is spatially dependent, leading to distinct cell motility in the core and periphery. In order to understand the  dynamical phase space structure of cells predicted by t-SNE, we probed the nature of forces exerted on the cells. We first calculated force-force persistence of the $i^{th}$ cell, $FF_i(t)=\frac{{\bf F}_i(t+\delta t)\cdot {\bf F}_i(t)}{|{\bf F}_i(t+\delta t)||{\bf F}_i(t)|}$ where ${\bf F}_i(t)$ ($|{\bf F}_i(t)|$) is the force (force magnitude) on the $i^{th}$ cell at time $t$ and $\delta t = 0.05~\tau$ or $40~mins$. $FF_i(t)$ is a measure of force persistence an $i^{th}$ cell experiences and takes on values [-1,1]. If $FF(t)=1$ ($FF(t)=-1$), a cell experiences force in the same (opposite) direction at time $t$ and $t+\delta t$. We calculated $FF_i(t)$, with $\tau\leq t \leq 11 \tau$, for all the cells that belong to core and periphery and performed t-SNE analysis.  The t-SNE projection of the $FF(t)$ data in  Figure \ref{force_cell}a reveals contrasting force persistence for cells in the core and periphery. $FF(t)$ in the two regions partition into two disjoint sets, which is vividly illustrated in Figure \ref{force_cell}a. The distinct behavior of force persistence is indicative of the super-diffusive and sub-diffusive behavior of cells in the periphery and core respectively \cite{valencia2015collective, kang2020tumor}. The contrasting force behavior in core and periphery is intrinsically related to spatial propensity for cell division. The increased stress (due to jamming) in the core suppresses cell division whereas cells on the periphery can readily divide \cite{Dolega17NC, alessandri2013cellular}. This imbalance in cell division in the two regions, leads to contrasting force persistence. 

We calculated the $\frac{t_c}{\tau}$ distribution to quantify how long SGAF is persistent in the two regions. In order to extract $\frac{t_c}{\tau}$, we fit $FF_i(t)$ using Gaussian process regression (GPR) with the standard RBF kernel ($k(t,t')\sim e^{-\frac{(t-t')^2}{2t_c^2}}$). Figure \ref{force_cell}b  shows that the distribution of $\frac{t_c}{\tau}$ for cells in the core and periphery are resolvable. Furthermore, we find  that the mean persistence time in the periphery ($\frac{t_c}{\tau}=0.08$) is greater than in the core ($\frac{t_c}{\tau}=0.06$). Increased persistence of forces in the periphery results in greater directed movement. Inset in figure \ref{force_cell}b show FF(t) fit using GPR for one cell in core and periphery. Experiments have noted that the cells in the periphery move predominantly radially outward \cite{valencia2015collective}. Therefore, we calculated the radial force $\frac{{\bf F_i}_r}{|{\bf F_i}|}$ exerted on the cells in the two regions. Here, ${\bf F}_{i,r}={\bf F}_i\cdot \hat{{\bf r}_i}$, $\hat{{\bf r}_i}$ is the radial unit vector, $\hat{{\bf r}_i}= \frac{{\bf r}_i-{\bf r}_{com}}{|{\bf r}_i-{\bf r}_{com}|}$, and ${\bf r}_{com}$ is the center of mass of the tumor. If $\frac{{\bf F_i}_r}{|{\bf F_i}|}=1$, the force is radially directed outward whereas $\frac{{\bf F_i}_r}{|{\bf F_i}|}=-1$ implies inwardly directed force.  The probability distribution of $\frac{{\bf F_i}_r}{|{\bf F_i}|}$ for cells in the core is predominantly skewed more towards unity (Figure \ref{force_cell}c) than the core cells. The radially outward force explains the invasive characteristics of the cells at the tumor boundary \cite{valencia2015collective}. These forces originate solely due to the imbalance in cell division in the core and peripheral region of the tumor. Cell division as a source of active stress has been reported before \cite{doostmohammadi2015celebrating}, but the emergence of highly persistent nature of forces adds new insights into the physics of tumor expansion.

\floatsetup[figure]{style=plain,subcapbesideposition=top}
\begin{figure}
\includegraphics[width=0.9\linewidth] {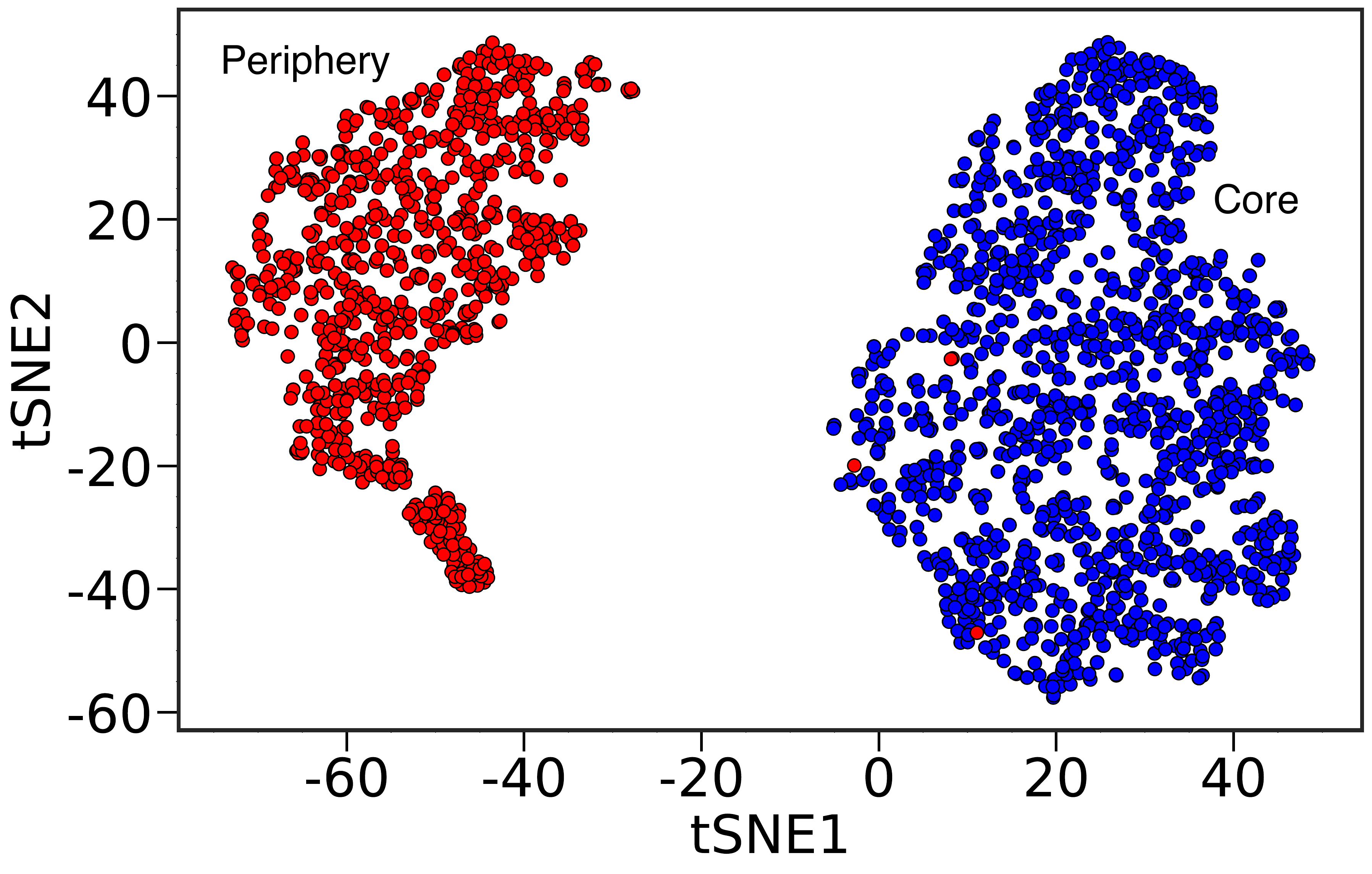}\label{penetration}
\caption{{\bf Core and Periphery cells have resolvable interpenetration}. Interpenetration, $h_{ij}$ (defined in text),  based classification for core and periphery cells using t-SNE. Cells in the core (periphery) are represented using blue (red) dots. The t-SNE algorithm resolves the interpenetration data for cells in the core and periphery. }
\label{interpenetration}
\end{figure}

Armed with the unbiased identification of the cells in the core and the periphery, we set out to find out if these features are manifested in other characteristics as well.  Experiments have established that the core cells are tightly packed or jammed \cite{puliafito2012collective}. Therefore, we expected that the inter-cellular distance can be used to differentiate between the cells in the two regions. We recorded the interpenetration ($h_{ij}$) distances ($h_{ij}$s) for cells in the time window $T_{w1}=\tau_{min}<t<11\tau_{min}$. Figure \ref{interpenetration} shows the result of t-SNE clustering based on $h_{ij}$ data for cells in the core and periphery ($\mathcal{P}=50$). To our surprise, the t-SNE algorithm clustered the cells remarkably well with clear phase separation between the cells in the core and the periphery. These results imply that the density in the two regions differ greatly, because the interior cells are jammed whereas the motility of the cells near the periphery is high. The emergence of the radially-dependent  density, with jamming in the interior,   is consistent with experiments that that show that pressure is higher in the core than at the periphery \cite{Dolega17NC}. 

\begin{figure}
\includegraphics[width=0.9\linewidth] {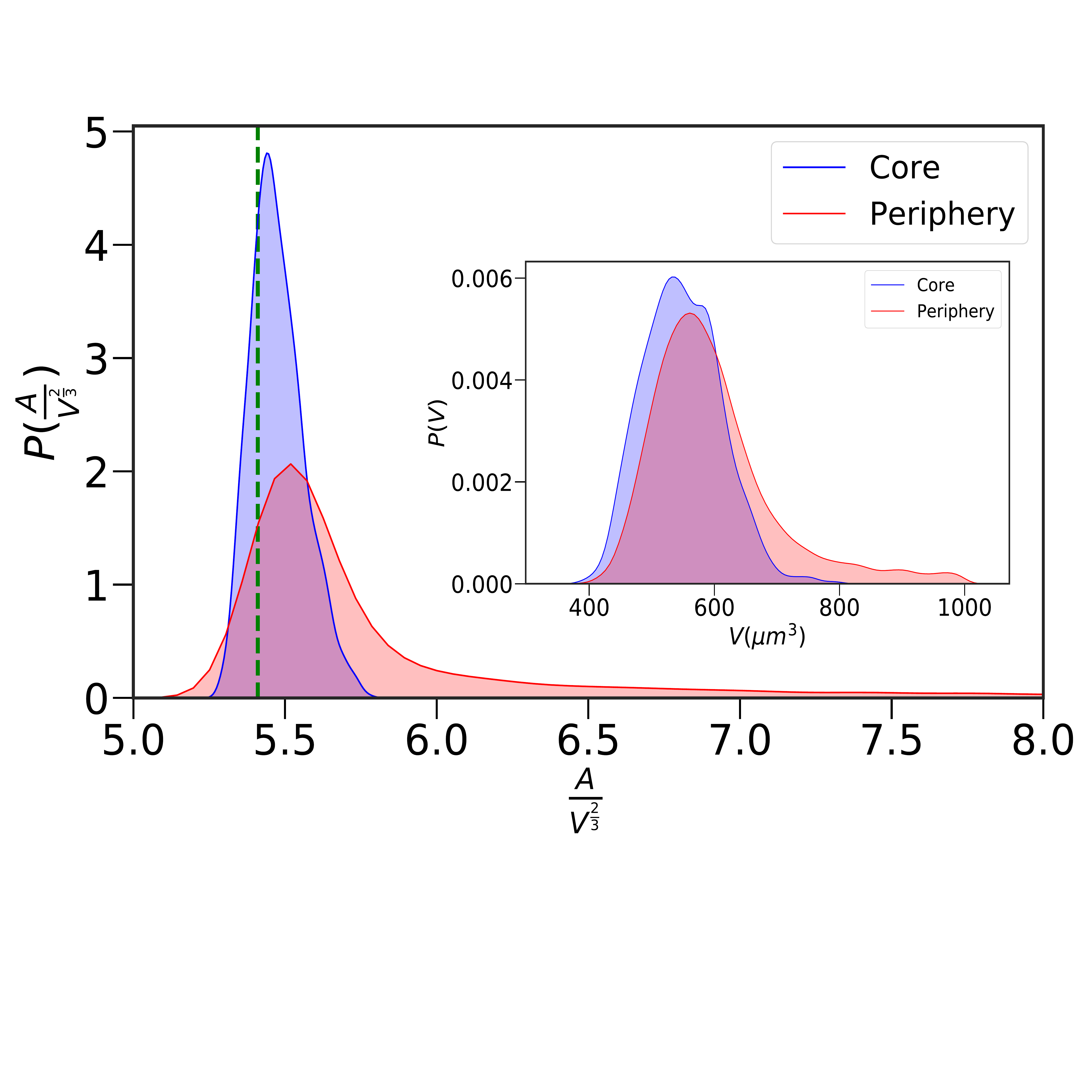}
\caption{{\bf Distribution of the shape parameter in the core and the periphery.} Blue curve shows the distribution for the cells in the core and red is for the cells in the periphery. The green line demarcates the solid to fluid boundary at $\frac{A}{V^{\frac{2}{3}}}=5.41$. The inset shows the distributions of the voronoi volume of cells in the core and the periphery.}
\label{shape_param}
\end{figure}

In order to provide a geometrical interpretation of the SGAF-driven phase separation, we followed Merkel and Manning  \cite{merkel2018geometrically} who predicted that  $S = \frac{A}{V^{\frac{2}{3}}}$ could serve as an order parameter for the rigidity transition in 3D confluent tissues. The variables $A$ and $V$  are, respectively,  the surface area and volume of the cell. The rigidity transition occurs at $S=5.41$ in three dimensions (see \cite{Bi16PRX} for results in two dimensions). Because the core (periphery) is solid-like (fluid-like), we expected that the shape parameter would reveal the observed differences in the motilities within a {\it single} tumor. We calculated the voronoi volume and area of the cells in at time $\approx 11 \tau$. We excluded the cells at the boundary as their voronoi volume is not defined. Figure \ref{shape_param} shows the distributions of the shape parameter distribution in the interior and the periphery.  Remarkably, the distribution for cells in the core is narrow with a peak at 5.41, close to the predicted \cite{merkel2018geometrically} solid to fluid transition value for confluent tissues.  However, the packing of cells in the interior in our simulations is qualitatively different, and does not reach confluency. In contrast, the distribution in the boundary are broadly peaked with a mean around 5.6. The inset in figure \ref{shape_param} shows that the cells in boundary have a bigger voronoi volume as compared to cells in the core.  We should emphasize the variations in the shape parameter is observed in a tumor in which the low motile and high motile cells are simultaneously present. It appears that the shape parameter is good predictor of the transition from a jammed to a motile (super diffusive) state even in a continuously growing tumor whose dynamics is determined by cell division and apoptosis. 

\floatsetup[figure]{style=plain,subcapbesideposition=top}


We used the unsupervised clustering technique (t-SNE)\cite{maaten2008visualizing,van2014accelerating,van2012visualizing} to elucidate the extent of heterogeneity in an evolving solid tumor consisting of nearly identical cells. The unbiased t-SNE analysis of the simulation data shows unambiguously that the dynamical behavior of cells in a growing tumor spheroid depends on the distance from the tumor core. The gradual change in the dynamical behavior from a jammed state in the tumor interior to highly motile (super-diffusive) behavior at the periphery is due to the generation of self-generated persistent forces that arises dynamically due to the inequality between cell division and apoptosis rates.   The t-SNE method resolves the dynamical phase space structure of identical cells, revealing a plausible mechanism for non-equilibrium phase separation  \cite{kang2020tumor}. Our results, establishing dynamic heterogeneity in a single tumor consisting of nearly identical cells, imply that  average properties in non-equilibrium systems may have little physical meaning \cite{altschuler2010cellular}.

 We would like to thank Xin Li and Mauro Mugnai for valuable comments on the manuscript. This work was supported by grants from National Science Foundation (PHY 17-08128, PHY-1522550). Additional support was provided by the Collie-Welch Reagents Chair (F-0019). 

\section*{DATA AVAILABILITY}
The data that support the findings of this study are available from the corresponding author upon reasonable request.

\vskip 0.2in
\bibliography{tsne}

\begin{thebibliography}{40}%
\makeatletter
\providecommand \@ifxundefined [1]{%
 \@ifx{#1\undefined}
}%
\providecommand \@ifnum [1]{%
 \ifnum #1\expandafter \@firstoftwo
 \else \expandafter \@secondoftwo
 \fi
}%
\providecommand \@ifx [1]{%
 \ifx #1\expandafter \@firstoftwo
 \else \expandafter \@secondoftwo
 \fi
}%
\providecommand \natexlab [1]{#1}%
\providecommand \enquote  [1]{``#1''}%
\providecommand \bibnamefont  [1]{#1}%
\providecommand \bibfnamefont [1]{#1}%
\providecommand \citenamefont [1]{#1}%
\providecommand \href@noop [0]{\@secondoftwo}%
\providecommand \href [0]{\begingroup \@sanitize@url \@href}%
\providecommand \@href[1]{\@@startlink{#1}\@@href}%
\providecommand \@@href[1]{\endgroup#1\@@endlink}%
\providecommand \@sanitize@url [0]{\catcode `\\12\catcode `\$12\catcode
  `\&12\catcode `\#12\catcode `\^12\catcode `\_12\catcode `\%12\relax}%
\providecommand \@@startlink[1]{}%
\providecommand \@@endlink[0]{}%
\providecommand \url  [0]{\begingroup\@sanitize@url \@url }%
\providecommand \@url [1]{\endgroup\@href {#1}{\urlprefix }}%
\providecommand \urlprefix  [0]{URL }%
\providecommand \Eprint [0]{\href }%
\providecommand \doibase [0]{http://dx.doi.org/}%
\providecommand \selectlanguage [0]{\@gobble}%
\providecommand \bibinfo  [0]{\@secondoftwo}%
\providecommand \bibfield  [0]{\@secondoftwo}%
\providecommand \translation [1]{[#1]}%
\providecommand \BibitemOpen [0]{}%
\providecommand \bibitemStop [0]{}%
\providecommand \bibitemNoStop [0]{.\EOS\space}%
\providecommand \EOS [0]{\spacefactor3000\relax}%
\providecommand \BibitemShut  [1]{\csname bibitem#1\endcsname}%
\let\auto@bib@innerbib\@empty
\bibitem [{\citenamefont {Heppner}, \citenamefont {Miller}\ \emph
  {et~al.}(1983)\citenamefont {Heppner}, \citenamefont {Miller} \emph
  {et~al.}}]{Heppner83CancerMetastasisRev}%
  \BibitemOpen
  \bibfield  {author} {\bibinfo {author} {\bibfnamefont {G.~H.}\ \bibnamefont
  {Heppner}}, \bibinfo {author} {\bibfnamefont {B.~E.}\ \bibnamefont {Miller}},
   \emph {et~al.},\ }\bibfield  {title} {\enquote {\bibinfo {title} {Tumor
  subpopulation interactions in neoplasms},}\ }\href@noop {} {\bibfield
  {journal} {\bibinfo  {journal} {Biochimica et Biophysica Acta (BBA)-Reviews
  on Cancer}\ }\textbf {\bibinfo {volume} {695}},\ \bibinfo {pages} {215--226}
  (\bibinfo {year} {1983})}\BibitemShut {NoStop}%
\bibitem [{\citenamefont {McGranahan}\ and\ \citenamefont
  {Swanton}(2015)}]{mcgranahan2015biological}%
  \BibitemOpen
  \bibfield  {author} {\bibinfo {author} {\bibfnamefont {N.}~\bibnamefont
  {McGranahan}}\ and\ \bibinfo {author} {\bibfnamefont {C.}~\bibnamefont
  {Swanton}},\ }\bibfield  {title} {\enquote {\bibinfo {title} {Biological and
  therapeutic impact of intratumor heterogeneity in cancer evolution},}\
  }\href@noop {} {\bibfield  {journal} {\bibinfo  {journal} {Cancer cell}\
  }\textbf {\bibinfo {volume} {27}},\ \bibinfo {pages} {15--26} (\bibinfo
  {year} {2015})}\BibitemShut {NoStop}%
\bibitem [{\citenamefont {Hinohara}\ and\ \citenamefont
  {Polyak}(2019)}]{hinohara2019intratumoral}%
  \BibitemOpen
  \bibfield  {author} {\bibinfo {author} {\bibfnamefont {K.}~\bibnamefont
  {Hinohara}}\ and\ \bibinfo {author} {\bibfnamefont {K.}~\bibnamefont
  {Polyak}},\ }\bibfield  {title} {\enquote {\bibinfo {title} {Intratumoral
  heterogeneity: more than just mutations},}\ }\href@noop {} {\bibfield
  {journal} {\bibinfo  {journal} {Trends in cell biology}\ } (\bibinfo {year}
  {2019})}\BibitemShut {NoStop}%
\bibitem [{\citenamefont {Gerlinger}\ \emph {et~al.}(2012)\citenamefont
  {Gerlinger}, \citenamefont {Rowan}, \citenamefont {Horswell}, \citenamefont
  {Larkin}, \citenamefont {Endesfelder}, \citenamefont {Gronroos},
  \citenamefont {Martinez}, \citenamefont {Matthews}, \citenamefont {Stewart},
  \citenamefont {Tarpey} \emph {et~al.}}]{gerlinger2012intratumor}%
  \BibitemOpen
  \bibfield  {author} {\bibinfo {author} {\bibfnamefont {M.}~\bibnamefont
  {Gerlinger}}, \bibinfo {author} {\bibfnamefont {A.~J.}\ \bibnamefont
  {Rowan}}, \bibinfo {author} {\bibfnamefont {S.}~\bibnamefont {Horswell}},
  \bibinfo {author} {\bibfnamefont {J.}~\bibnamefont {Larkin}}, \bibinfo
  {author} {\bibfnamefont {D.}~\bibnamefont {Endesfelder}}, \bibinfo {author}
  {\bibfnamefont {E.}~\bibnamefont {Gronroos}}, \bibinfo {author}
  {\bibfnamefont {P.}~\bibnamefont {Martinez}}, \bibinfo {author}
  {\bibfnamefont {N.}~\bibnamefont {Matthews}}, \bibinfo {author}
  {\bibfnamefont {A.}~\bibnamefont {Stewart}}, \bibinfo {author} {\bibfnamefont
  {P.}~\bibnamefont {Tarpey}},  \emph {et~al.},\ }\bibfield  {title} {\enquote
  {\bibinfo {title} {Intratumor heterogeneity and branched evolution revealed
  by multiregion sequencing},}\ }\href@noop {} {\bibfield  {journal} {\bibinfo
  {journal} {New England journal of medicine}\ }\textbf {\bibinfo {volume}
  {366}},\ \bibinfo {pages} {883--892} (\bibinfo {year} {2012})}\BibitemShut
  {NoStop}%
\bibitem [{\citenamefont {Gerlinger}\ \emph {et~al.}(2014)\citenamefont
  {Gerlinger}, \citenamefont {Horswell}, \citenamefont {Larkin}, \citenamefont
  {Rowan}, \citenamefont {Salm}, \citenamefont {Varela}, \citenamefont
  {Fisher}, \citenamefont {McGranahan}, \citenamefont {Matthews}, \citenamefont
  {Santos} \emph {et~al.}}]{gerlinger2014}%
  \BibitemOpen
  \bibfield  {author} {\bibinfo {author} {\bibfnamefont {M.}~\bibnamefont
  {Gerlinger}}, \bibinfo {author} {\bibfnamefont {S.}~\bibnamefont {Horswell}},
  \bibinfo {author} {\bibfnamefont {J.}~\bibnamefont {Larkin}}, \bibinfo
  {author} {\bibfnamefont {A.~J.}\ \bibnamefont {Rowan}}, \bibinfo {author}
  {\bibfnamefont {M.~P.}\ \bibnamefont {Salm}}, \bibinfo {author}
  {\bibfnamefont {I.}~\bibnamefont {Varela}}, \bibinfo {author} {\bibfnamefont
  {R.}~\bibnamefont {Fisher}}, \bibinfo {author} {\bibfnamefont
  {N.}~\bibnamefont {McGranahan}}, \bibinfo {author} {\bibfnamefont
  {N.}~\bibnamefont {Matthews}}, \bibinfo {author} {\bibfnamefont {C.~R.}\
  \bibnamefont {Santos}},  \emph {et~al.},\ }\bibfield  {title} {\enquote
  {\bibinfo {title} {Genomic architecture and evolution of clear cell renal
  cell carcinomas defined by multiregion sequencing},}\ }\href@noop {}
  {\bibfield  {journal} {\bibinfo  {journal} {Nature genetics}\ }\textbf
  {\bibinfo {volume} {46}},\ \bibinfo {pages} {225} (\bibinfo {year}
  {2014})}\BibitemShut {NoStop}%
\bibitem [{\citenamefont {Harbst}\ \emph {et~al.}(2016)\citenamefont {Harbst},
  \citenamefont {Lauss}, \citenamefont {Cirenajwis}, \citenamefont {Isaksson},
  \citenamefont {Rosengren}, \citenamefont {Torngren}, \citenamefont {Kvist},
  \citenamefont {Johansson}, \citenamefont {Vallon-Christersson}, \citenamefont
  {Baldetorp} \emph {et~al.}}]{harbst2016}%
  \BibitemOpen
  \bibfield  {author} {\bibinfo {author} {\bibfnamefont {K.}~\bibnamefont
  {Harbst}}, \bibinfo {author} {\bibfnamefont {M.}~\bibnamefont {Lauss}},
  \bibinfo {author} {\bibfnamefont {H.}~\bibnamefont {Cirenajwis}}, \bibinfo
  {author} {\bibfnamefont {K.}~\bibnamefont {Isaksson}}, \bibinfo {author}
  {\bibfnamefont {F.}~\bibnamefont {Rosengren}}, \bibinfo {author}
  {\bibfnamefont {T.}~\bibnamefont {Torngren}}, \bibinfo {author}
  {\bibfnamefont {A.}~\bibnamefont {Kvist}}, \bibinfo {author} {\bibfnamefont
  {M.~C.}\ \bibnamefont {Johansson}}, \bibinfo {author} {\bibfnamefont
  {J.}~\bibnamefont {Vallon-Christersson}}, \bibinfo {author} {\bibfnamefont
  {B.}~\bibnamefont {Baldetorp}},  \emph {et~al.},\ }\bibfield  {title}
  {\enquote {\bibinfo {title} {Multi-region whole-exome sequencing uncovers the
  genetic evolution and mutational heterogeneity of early-stage metastatic
  melanoma},}\ }\href@noop {} {\bibfield  {journal} {\bibinfo  {journal}
  {Cancer research}\ ,\ \bibinfo {pages} {canres--3476}} (\bibinfo {year}
  {2016})}\BibitemShut {NoStop}%
\bibitem [{\citenamefont {de~Bruin}\ \emph {et~al.}(2014)\citenamefont
  {de~Bruin}, \citenamefont {McGranahan}, \citenamefont {Mitter}, \citenamefont
  {Salm}, \citenamefont {Wedge}, \citenamefont {Yates}, \citenamefont
  {Jamal-Hanjani}, \citenamefont {Shafi}, \citenamefont {Murugaesu},
  \citenamefont {Rowan} \emph {et~al.}}]{debruin2014}%
  \BibitemOpen
  \bibfield  {author} {\bibinfo {author} {\bibfnamefont {E.~C.}\ \bibnamefont
  {de~Bruin}}, \bibinfo {author} {\bibfnamefont {N.}~\bibnamefont
  {McGranahan}}, \bibinfo {author} {\bibfnamefont {R.}~\bibnamefont {Mitter}},
  \bibinfo {author} {\bibfnamefont {M.}~\bibnamefont {Salm}}, \bibinfo {author}
  {\bibfnamefont {D.~C.}\ \bibnamefont {Wedge}}, \bibinfo {author}
  {\bibfnamefont {L.}~\bibnamefont {Yates}}, \bibinfo {author} {\bibfnamefont
  {M.}~\bibnamefont {Jamal-Hanjani}}, \bibinfo {author} {\bibfnamefont
  {S.}~\bibnamefont {Shafi}}, \bibinfo {author} {\bibfnamefont
  {N.}~\bibnamefont {Murugaesu}}, \bibinfo {author} {\bibfnamefont {A.~J.}\
  \bibnamefont {Rowan}},  \emph {et~al.},\ }\bibfield  {title} {\enquote
  {\bibinfo {title} {Spatial and temporal diversity in genomic instability
  processes defines lung cancer evolution},}\ }\href@noop {} {\bibfield
  {journal} {\bibinfo  {journal} {Science}\ }\textbf {\bibinfo {volume}
  {346}},\ \bibinfo {pages} {251--256} (\bibinfo {year} {2014})}\BibitemShut
  {NoStop}%
\bibitem [{\citenamefont {Zhang}\ \emph {et~al.}(2014)\citenamefont {Zhang},
  \citenamefont {Fujimoto}, \citenamefont {Zhang}, \citenamefont {Wedge},
  \citenamefont {Song}, \citenamefont {Zhang}, \citenamefont {Seth},
  \citenamefont {Chow}, \citenamefont {Cao}, \citenamefont {Gumbs} \emph
  {et~al.}}]{zhang2014intratumor}%
  \BibitemOpen
  \bibfield  {author} {\bibinfo {author} {\bibfnamefont {J.}~\bibnamefont
  {Zhang}}, \bibinfo {author} {\bibfnamefont {J.}~\bibnamefont {Fujimoto}},
  \bibinfo {author} {\bibfnamefont {J.}~\bibnamefont {Zhang}}, \bibinfo
  {author} {\bibfnamefont {D.~C.}\ \bibnamefont {Wedge}}, \bibinfo {author}
  {\bibfnamefont {X.}~\bibnamefont {Song}}, \bibinfo {author} {\bibfnamefont
  {J.}~\bibnamefont {Zhang}}, \bibinfo {author} {\bibfnamefont
  {S.}~\bibnamefont {Seth}}, \bibinfo {author} {\bibfnamefont {C.-W.}\
  \bibnamefont {Chow}}, \bibinfo {author} {\bibfnamefont {Y.}~\bibnamefont
  {Cao}}, \bibinfo {author} {\bibfnamefont {C.}~\bibnamefont {Gumbs}},  \emph
  {et~al.},\ }\bibfield  {title} {\enquote {\bibinfo {title} {Intratumor
  heterogeneity in localized lung adenocarcinomas delineated by multiregion
  sequencing},}\ }\href@noop {} {\bibfield  {journal} {\bibinfo  {journal}
  {Science}\ }\textbf {\bibinfo {volume} {346}},\ \bibinfo {pages} {256--259}
  (\bibinfo {year} {2014})}\BibitemShut {NoStop}%
\bibitem [{\citenamefont {Cao}\ \emph {et~al.}(2015)\citenamefont {Cao},
  \citenamefont {Wu}, \citenamefont {Yan}, \citenamefont {Tian}, \citenamefont
  {Ma}, \citenamefont {Zhang}, \citenamefont {Li}, \citenamefont {Han},
  \citenamefont {Liu}, \citenamefont {Gu} \emph {et~al.}}]{cao2015}%
  \BibitemOpen
  \bibfield  {author} {\bibinfo {author} {\bibfnamefont {W.}~\bibnamefont
  {Cao}}, \bibinfo {author} {\bibfnamefont {W.}~\bibnamefont {Wu}}, \bibinfo
  {author} {\bibfnamefont {M.}~\bibnamefont {Yan}}, \bibinfo {author}
  {\bibfnamefont {F.}~\bibnamefont {Tian}}, \bibinfo {author} {\bibfnamefont
  {C.}~\bibnamefont {Ma}}, \bibinfo {author} {\bibfnamefont {Q.}~\bibnamefont
  {Zhang}}, \bibinfo {author} {\bibfnamefont {X.}~\bibnamefont {Li}}, \bibinfo
  {author} {\bibfnamefont {P.}~\bibnamefont {Han}}, \bibinfo {author}
  {\bibfnamefont {Z.}~\bibnamefont {Liu}}, \bibinfo {author} {\bibfnamefont
  {J.}~\bibnamefont {Gu}},  \emph {et~al.},\ }\bibfield  {title} {\enquote
  {\bibinfo {title} {Multiple region whole-exome sequencing reveals
  dramatically evolving intratumor genomic heterogeneity in esophageal squamous
  cell carcinoma},}\ }\href@noop {} {\bibfield  {journal} {\bibinfo  {journal}
  {Oncogenesis}\ }\textbf {\bibinfo {volume} {4}},\ \bibinfo {pages} {e175}
  (\bibinfo {year} {2015})}\BibitemShut {NoStop}%
\bibitem [{\citenamefont {Richards}\ \emph {et~al.}(2018)\citenamefont
  {Richards}, \citenamefont {Mason}, \citenamefont {Levy}, \citenamefont
  {Bearon},\ and\ \citenamefont {See}}]{richards20184d}%
  \BibitemOpen
  \bibfield  {author} {\bibinfo {author} {\bibfnamefont {R.}~\bibnamefont
  {Richards}}, \bibinfo {author} {\bibfnamefont {D.}~\bibnamefont {Mason}},
  \bibinfo {author} {\bibfnamefont {R.}~\bibnamefont {Levy}}, \bibinfo {author}
  {\bibfnamefont {R.}~\bibnamefont {Bearon}}, \ and\ \bibinfo {author}
  {\bibfnamefont {V.}~\bibnamefont {See}},\ }\bibfield  {title} {\enquote
  {\bibinfo {title} {4d imaging and analysis of multicellular tumour spheroid
  cell migration and invasion},}\ }\href@noop {} {\bibfield  {journal}
  {\bibinfo  {journal} {bioRxiv}\ ,\ \bibinfo {pages} {443648}} (\bibinfo
  {year} {2018})}\BibitemShut {NoStop}%
\bibitem [{\citenamefont {Valencia}\ \emph {et~al.}(2015)\citenamefont
  {Valencia}, \citenamefont {Wu}, \citenamefont {Yogurtcu}, \citenamefont
  {Rao}, \citenamefont {DiGiacomo}, \citenamefont {Godet}, \citenamefont {He},
  \citenamefont {Lee}, \citenamefont {Gilkes}, \citenamefont {Sun} \emph
  {et~al.}}]{valencia2015collective}%
  \BibitemOpen
  \bibfield  {author} {\bibinfo {author} {\bibfnamefont {A.~M.~J.}\
  \bibnamefont {Valencia}}, \bibinfo {author} {\bibfnamefont {P.-H.}\
  \bibnamefont {Wu}}, \bibinfo {author} {\bibfnamefont {O.~N.}\ \bibnamefont
  {Yogurtcu}}, \bibinfo {author} {\bibfnamefont {P.}~\bibnamefont {Rao}},
  \bibinfo {author} {\bibfnamefont {J.}~\bibnamefont {DiGiacomo}}, \bibinfo
  {author} {\bibfnamefont {I.}~\bibnamefont {Godet}}, \bibinfo {author}
  {\bibfnamefont {L.}~\bibnamefont {He}}, \bibinfo {author} {\bibfnamefont
  {M.-H.}\ \bibnamefont {Lee}}, \bibinfo {author} {\bibfnamefont
  {D.}~\bibnamefont {Gilkes}}, \bibinfo {author} {\bibfnamefont {S.~X.}\
  \bibnamefont {Sun}},  \emph {et~al.},\ }\bibfield  {title} {\enquote
  {\bibinfo {title} {Collective cancer cell invasion induced by coordinated
  contractile stresses},}\ }\href@noop {} {\bibfield  {journal} {\bibinfo
  {journal} {Oncotarget}\ }\textbf {\bibinfo {volume} {6}},\ \bibinfo {pages}
  {43438} (\bibinfo {year} {2015})}\BibitemShut {NoStop}%
\bibitem [{\citenamefont {Kang}\ \emph {et~al.}(2020)\citenamefont {Kang},
  \citenamefont {Ferruzzi}, \citenamefont {Spatarelu}, \citenamefont {Han},
  \citenamefont {Sharma}, \citenamefont {Koehler}, \citenamefont {Butler},
  \citenamefont {Roblyer}, \citenamefont {Zaman}, \citenamefont {Guo} \emph
  {et~al.}}]{kang2020tumor}%
  \BibitemOpen
  \bibfield  {author} {\bibinfo {author} {\bibfnamefont {W.}~\bibnamefont
  {Kang}}, \bibinfo {author} {\bibfnamefont {J.}~\bibnamefont {Ferruzzi}},
  \bibinfo {author} {\bibfnamefont {C.-P.}\ \bibnamefont {Spatarelu}}, \bibinfo
  {author} {\bibfnamefont {Y.~L.}\ \bibnamefont {Han}}, \bibinfo {author}
  {\bibfnamefont {Y.}~\bibnamefont {Sharma}}, \bibinfo {author} {\bibfnamefont
  {S.}~\bibnamefont {Koehler}}, \bibinfo {author} {\bibfnamefont {J.~P.}\
  \bibnamefont {Butler}}, \bibinfo {author} {\bibfnamefont {D.}~\bibnamefont
  {Roblyer}}, \bibinfo {author} {\bibfnamefont {M.~H.}\ \bibnamefont {Zaman}},
  \bibinfo {author} {\bibfnamefont {M.}~\bibnamefont {Guo}},  \emph {et~al.},\
  }\bibfield  {title} {\enquote {\bibinfo {title} {Tumor invasion as
  non-equilibrium phase separation},}\ }\href@noop {} {\bibfield  {journal}
  {\bibinfo  {journal} {bioRxiv}\ } (\bibinfo {year} {2020})}\BibitemShut
  {NoStop}%
\bibitem [{\citenamefont {Han}\ \emph {et~al.}(2019)\citenamefont {Han},
  \citenamefont {Pegoraro}, \citenamefont {Li}, \citenamefont {Li},
  \citenamefont {Yuan}, \citenamefont {Xu}, \citenamefont {Gu}, \citenamefont
  {Sun}, \citenamefont {Hao}, \citenamefont {Gupta} \emph
  {et~al.}}]{han2019cell}%
  \BibitemOpen
  \bibfield  {author} {\bibinfo {author} {\bibfnamefont {Y.~L.}\ \bibnamefont
  {Han}}, \bibinfo {author} {\bibfnamefont {A.~F.}\ \bibnamefont {Pegoraro}},
  \bibinfo {author} {\bibfnamefont {H.}~\bibnamefont {Li}}, \bibinfo {author}
  {\bibfnamefont {K.}~\bibnamefont {Li}}, \bibinfo {author} {\bibfnamefont
  {Y.}~\bibnamefont {Yuan}}, \bibinfo {author} {\bibfnamefont {G.}~\bibnamefont
  {Xu}}, \bibinfo {author} {\bibfnamefont {Z.}~\bibnamefont {Gu}}, \bibinfo
  {author} {\bibfnamefont {J.}~\bibnamefont {Sun}}, \bibinfo {author}
  {\bibfnamefont {Y.}~\bibnamefont {Hao}}, \bibinfo {author} {\bibfnamefont
  {S.~K.}\ \bibnamefont {Gupta}},  \emph {et~al.},\ }\bibfield  {title}
  {\enquote {\bibinfo {title} {Cell swelling, softening and invasion in a
  three-dimensional breast cancer model},}\ }\href@noop {} {\bibfield
  {journal} {\bibinfo  {journal} {Nature Physics}\ ,\ \bibinfo {pages} {1--8}}
  (\bibinfo {year} {2019})}\BibitemShut {NoStop}%
\bibitem [{\citenamefont {Palamidessi}\ \emph {et~al.}(2019)\citenamefont
  {Palamidessi}, \citenamefont {Malinverno}, \citenamefont {Frittoli},
  \citenamefont {Corallino}, \citenamefont {Barbieri}, \citenamefont
  {Sigismund}, \citenamefont {Beznoussenko}, \citenamefont {Martini},
  \citenamefont {Garre}, \citenamefont {Ferrara}, \citenamefont {Tripodo},
  \citenamefont {Ascione}, \citenamefont {Cavalcanti-Adam}, \citenamefont {Li},
  \citenamefont {Di~Fiore}, \citenamefont {Parazzoli}, \citenamefont
  {Giavazzi}, \citenamefont {Cerbino},\ and\ \citenamefont
  {Scita}}]{palamidessi2019unjamming}%
  \BibitemOpen
  \bibfield  {author} {\bibinfo {author} {\bibfnamefont {A.}~\bibnamefont
  {Palamidessi}}, \bibinfo {author} {\bibfnamefont {C.}~\bibnamefont
  {Malinverno}}, \bibinfo {author} {\bibfnamefont {E.}~\bibnamefont
  {Frittoli}}, \bibinfo {author} {\bibfnamefont {S.}~\bibnamefont {Corallino}},
  \bibinfo {author} {\bibfnamefont {E.}~\bibnamefont {Barbieri}}, \bibinfo
  {author} {\bibfnamefont {S.}~\bibnamefont {Sigismund}}, \bibinfo {author}
  {\bibfnamefont {G.~V.}\ \bibnamefont {Beznoussenko}}, \bibinfo {author}
  {\bibfnamefont {E.}~\bibnamefont {Martini}}, \bibinfo {author} {\bibfnamefont
  {M.}~\bibnamefont {Garre}}, \bibinfo {author} {\bibfnamefont
  {I.}~\bibnamefont {Ferrara}}, \bibinfo {author} {\bibfnamefont
  {C.}~\bibnamefont {Tripodo}}, \bibinfo {author} {\bibfnamefont
  {F.}~\bibnamefont {Ascione}}, \bibinfo {author} {\bibfnamefont {E.~A.}\
  \bibnamefont {Cavalcanti-Adam}}, \bibinfo {author} {\bibfnamefont
  {Q.}~\bibnamefont {Li}}, \bibinfo {author} {\bibfnamefont {P.~P.}\
  \bibnamefont {Di~Fiore}}, \bibinfo {author} {\bibfnamefont {D.}~\bibnamefont
  {Parazzoli}}, \bibinfo {author} {\bibfnamefont {F.}~\bibnamefont {Giavazzi}},
  \bibinfo {author} {\bibfnamefont {R.}~\bibnamefont {Cerbino}}, \ and\
  \bibinfo {author} {\bibfnamefont {G.}~\bibnamefont {Scita}},\ }\bibfield
  {title} {\enquote {\bibinfo {title} {Unjamming overcomes kinetic and
  proliferation arrest in terminally differentiated cells and promotes
  collective motility of carcinoma},}\ }\href@noop {} {\bibfield  {journal}
  {\bibinfo  {journal} {Nature Materials}\ }\textbf {\bibinfo {volume} {18}},\
  \bibinfo {pages} {1252--1263} (\bibinfo {year} {2019})}\BibitemShut {NoStop}%
\bibitem [{\citenamefont {Sinha}\ \emph {et~al.}(2020)\citenamefont {Sinha},
  \citenamefont {Malmi-Kakkada}, \citenamefont {Li}, \citenamefont {Samanta},\
  and\ \citenamefont {Thirumalai}}]{sinha2020spatially}%
  \BibitemOpen
  \bibfield  {author} {\bibinfo {author} {\bibfnamefont {S.}~\bibnamefont
  {Sinha}}, \bibinfo {author} {\bibfnamefont {A.~N.}\ \bibnamefont
  {Malmi-Kakkada}}, \bibinfo {author} {\bibfnamefont {X.}~\bibnamefont {Li}},
  \bibinfo {author} {\bibfnamefont {H.~S.}\ \bibnamefont {Samanta}}, \ and\
  \bibinfo {author} {\bibfnamefont {D.}~\bibnamefont {Thirumalai}},\ }\bibfield
   {title} {\enquote {\bibinfo {title} {Spatially heterogeneous dynamics of
  cells in a growing tumor spheroid: Comparison between theory and
  experiments},}\ }\href@noop {} {\bibfield  {journal} {\bibinfo  {journal}
  {Soft Matter}\ }\textbf {\bibinfo {volume} {16}},\ \bibinfo {pages}
  {5294--5304} (\bibinfo {year} {2020})}\BibitemShut {NoStop}%
\bibitem [{\citenamefont {Kirkpatrick}\ and\ \citenamefont
  {Thirumalai}(2015)}]{kirkpatrick2015colloquium}%
  \BibitemOpen
  \bibfield  {author} {\bibinfo {author} {\bibfnamefont {T.}~\bibnamefont
  {Kirkpatrick}}\ and\ \bibinfo {author} {\bibfnamefont {D.}~\bibnamefont
  {Thirumalai}},\ }\bibfield  {title} {\enquote {\bibinfo {title} {Colloquium:
  Random first order transition theory concepts in biology and physics},}\
  }\href@noop {} {\bibfield  {journal} {\bibinfo  {journal} {Reviews of Modern
  Physics}\ }\textbf {\bibinfo {volume} {87}},\ \bibinfo {pages} {183}
  (\bibinfo {year} {2015})}\BibitemShut {NoStop}%
\bibitem [{\citenamefont {Berthier}\ and\ \citenamefont
  {Barrat}(2002)}]{Berthier20JCP}%
  \BibitemOpen
  \bibfield  {author} {\bibinfo {author} {\bibfnamefont {L.}~\bibnamefont
  {Berthier}}\ and\ \bibinfo {author} {\bibfnamefont {J.-L.}\ \bibnamefont
  {Barrat}},\ }\bibfield  {title} {\enquote {\bibinfo {title} {Nonequilibrium
  dynamics and fluctuation-dissipation relation in a sheared fluid},}\
  }\href@noop {} {\bibfield  {journal} {\bibinfo  {journal} {The Journal of
  Chemical Physics}\ }\textbf {\bibinfo {volume} {116}},\ \bibinfo {pages}
  {6228--6242} (\bibinfo {year} {2002})}\BibitemShut {NoStop}%
\bibitem [{\citenamefont {Malmi-Kakkada}\ \emph {et~al.}(2018)\citenamefont
  {Malmi-Kakkada}, \citenamefont {Li}, \citenamefont {Samanta}, \citenamefont
  {Sinha},\ and\ \citenamefont {Thirumalai}}]{malmi2018cell}%
  \BibitemOpen
  \bibfield  {author} {\bibinfo {author} {\bibfnamefont {A.~N.}\ \bibnamefont
  {Malmi-Kakkada}}, \bibinfo {author} {\bibfnamefont {X.}~\bibnamefont {Li}},
  \bibinfo {author} {\bibfnamefont {H.~S.}\ \bibnamefont {Samanta}}, \bibinfo
  {author} {\bibfnamefont {S.}~\bibnamefont {Sinha}}, \ and\ \bibinfo {author}
  {\bibfnamefont {D.}~\bibnamefont {Thirumalai}},\ }\bibfield  {title}
  {\enquote {\bibinfo {title} {Cell growth rate dictates the onset of glass to
  fluidlike transition and long time superdiffusion in an evolving cell
  colony},}\ }\href@noop {} {\bibfield  {journal} {\bibinfo  {journal}
  {Physical Review X}\ }\textbf {\bibinfo {volume} {8}},\ \bibinfo {pages}
  {021025} (\bibinfo {year} {2018})}\BibitemShut {NoStop}%
\bibitem [{\citenamefont {Maaten}\ and\ \citenamefont
  {Hinton}(2008)}]{maaten2008visualizing}%
  \BibitemOpen
  \bibfield  {author} {\bibinfo {author} {\bibfnamefont {L.~v.~d.}\
  \bibnamefont {Maaten}}\ and\ \bibinfo {author} {\bibfnamefont
  {G.}~\bibnamefont {Hinton}},\ }\bibfield  {title} {\enquote {\bibinfo {title}
  {Visualizing data using t-sne},}\ }\href@noop {} {\bibfield  {journal}
  {\bibinfo  {journal} {Journal of machine learning research}\ }\textbf
  {\bibinfo {volume} {9}},\ \bibinfo {pages} {2579--2605} (\bibinfo {year}
  {2008})}\BibitemShut {NoStop}%
\bibitem [{\citenamefont {Van Der~Maaten}(2009)}]{van2009learning}%
  \BibitemOpen
  \bibfield  {author} {\bibinfo {author} {\bibfnamefont {L.}~\bibnamefont {Van
  Der~Maaten}},\ }\bibfield  {title} {\enquote {\bibinfo {title} {Learning a
  parametric embedding by preserving local structure},}\ }in\ \href@noop {}
  {\emph {\bibinfo {booktitle} {Artificial Intelligence and Statistics}}}\
  (\bibinfo {year} {2009})\ pp.\ \bibinfo {pages} {384--391}\BibitemShut
  {NoStop}%
\bibitem [{\citenamefont {Van~der Maaten}\ and\ \citenamefont
  {Hinton}(2012)}]{van2012visualizing}%
  \BibitemOpen
  \bibfield  {author} {\bibinfo {author} {\bibfnamefont {L.}~\bibnamefont
  {Van~der Maaten}}\ and\ \bibinfo {author} {\bibfnamefont {G.}~\bibnamefont
  {Hinton}},\ }\bibfield  {title} {\enquote {\bibinfo {title} {Visualizing
  non-metric similarities in multiple maps},}\ }\href@noop {} {\bibfield
  {journal} {\bibinfo  {journal} {Machine learning}\ }\textbf {\bibinfo
  {volume} {87}},\ \bibinfo {pages} {33--55} (\bibinfo {year}
  {2012})}\BibitemShut {NoStop}%
\bibitem [{\citenamefont {Van Der~Maaten}(2014)}]{van2014accelerating}%
  \BibitemOpen
  \bibfield  {author} {\bibinfo {author} {\bibfnamefont {L.}~\bibnamefont {Van
  Der~Maaten}},\ }\bibfield  {title} {\enquote {\bibinfo {title} {Accelerating
  t-sne using tree-based algorithms},}\ }\href@noop {} {\bibfield  {journal}
  {\bibinfo  {journal} {The Journal of Machine Learning Research}\ }\textbf
  {\bibinfo {volume} {15}},\ \bibinfo {pages} {3221--3245} (\bibinfo {year}
  {2014})}\BibitemShut {NoStop}%
\bibitem [{\citenamefont {Kobak}\ and\ \citenamefont
  {Berens}(2019)}]{kobak2019art}%
  \BibitemOpen
  \bibfield  {author} {\bibinfo {author} {\bibfnamefont {D.}~\bibnamefont
  {Kobak}}\ and\ \bibinfo {author} {\bibfnamefont {P.}~\bibnamefont {Berens}},\
  }\bibfield  {title} {\enquote {\bibinfo {title} {The art of using t-sne for
  single-cell transcriptomics},}\ }\href@noop {} {\bibfield  {journal}
  {\bibinfo  {journal} {Nature communications}\ }\textbf {\bibinfo {volume}
  {10}},\ \bibinfo {pages} {1--14} (\bibinfo {year} {2019})}\BibitemShut
  {NoStop}%
\bibitem [{\citenamefont {Linderman}\ \emph {et~al.}(2019)\citenamefont
  {Linderman}, \citenamefont {Rachh}, \citenamefont {Hoskins}, \citenamefont
  {Steinerberger},\ and\ \citenamefont {Kluger}}]{linderman2019fast}%
  \BibitemOpen
  \bibfield  {author} {\bibinfo {author} {\bibfnamefont {G.~C.}\ \bibnamefont
  {Linderman}}, \bibinfo {author} {\bibfnamefont {M.}~\bibnamefont {Rachh}},
  \bibinfo {author} {\bibfnamefont {J.~G.}\ \bibnamefont {Hoskins}}, \bibinfo
  {author} {\bibfnamefont {S.}~\bibnamefont {Steinerberger}}, \ and\ \bibinfo
  {author} {\bibfnamefont {Y.}~\bibnamefont {Kluger}},\ }\bibfield  {title}
  {\enquote {\bibinfo {title} {Fast interpolation-based t-sne for improved
  visualization of single-cell rna-seq data},}\ }\href@noop {} {\bibfield
  {journal} {\bibinfo  {journal} {Nature methods}\ }\textbf {\bibinfo {volume}
  {16}},\ \bibinfo {pages} {243--245} (\bibinfo {year} {2019})}\BibitemShut
  {NoStop}%
\bibitem [{\citenamefont {Dimitriadis}, \citenamefont {Neto},\ and\
  \citenamefont {Kampff}(2018)}]{dimitriadis2018t}%
  \BibitemOpen
  \bibfield  {author} {\bibinfo {author} {\bibfnamefont {G.}~\bibnamefont
  {Dimitriadis}}, \bibinfo {author} {\bibfnamefont {J.~P.}\ \bibnamefont
  {Neto}}, \ and\ \bibinfo {author} {\bibfnamefont {A.~R.}\ \bibnamefont
  {Kampff}},\ }\bibfield  {title} {\enquote {\bibinfo {title} {T-sne
  visualization of large-scale neural recordings},}\ }\href@noop {} {\bibfield
  {journal} {\bibinfo  {journal} {Neural computation}\ }\textbf {\bibinfo
  {volume} {30}},\ \bibinfo {pages} {1750--1774} (\bibinfo {year}
  {2018})}\BibitemShut {NoStop}%
\bibitem [{\citenamefont {Carrasquilla}\ and\ \citenamefont
  {Melko}(2017)}]{carrasquilla2017machine}%
  \BibitemOpen
  \bibfield  {author} {\bibinfo {author} {\bibfnamefont {J.}~\bibnamefont
  {Carrasquilla}}\ and\ \bibinfo {author} {\bibfnamefont {R.~G.}\ \bibnamefont
  {Melko}},\ }\bibfield  {title} {\enquote {\bibinfo {title} {Machine learning
  phases of matter},}\ }\href@noop {} {\bibfield  {journal} {\bibinfo
  {journal} {Nature Physics}\ }\textbf {\bibinfo {volume} {13}},\ \bibinfo
  {pages} {431--434} (\bibinfo {year} {2017})}\BibitemShut {NoStop}%
\bibitem [{\citenamefont {Ch'ng}, \citenamefont {Vazquez},\ and\ \citenamefont
  {Khatami}(2018)}]{ch2018unsupervised}%
  \BibitemOpen
  \bibfield  {author} {\bibinfo {author} {\bibfnamefont {K.}~\bibnamefont
  {Ch'ng}}, \bibinfo {author} {\bibfnamefont {N.}~\bibnamefont {Vazquez}}, \
  and\ \bibinfo {author} {\bibfnamefont {E.}~\bibnamefont {Khatami}},\
  }\bibfield  {title} {\enquote {\bibinfo {title} {Unsupervised machine
  learning account of magnetic transitions in the hubbard model},}\ }\href@noop
  {} {\bibfield  {journal} {\bibinfo  {journal} {Physical Review E}\ }\textbf
  {\bibinfo {volume} {97}},\ \bibinfo {pages} {013306} (\bibinfo {year}
  {2018})}\BibitemShut {NoStop}%
\bibitem [{\citenamefont {Day}\ \emph {et~al.}(2019)\citenamefont {Day},
  \citenamefont {Bukov}, \citenamefont {Weinberg}, \citenamefont {Mehta},\ and\
  \citenamefont {Sels}}]{day2019glassy}%
  \BibitemOpen
  \bibfield  {author} {\bibinfo {author} {\bibfnamefont {A.~G.}\ \bibnamefont
  {Day}}, \bibinfo {author} {\bibfnamefont {M.}~\bibnamefont {Bukov}}, \bibinfo
  {author} {\bibfnamefont {P.}~\bibnamefont {Weinberg}}, \bibinfo {author}
  {\bibfnamefont {P.}~\bibnamefont {Mehta}}, \ and\ \bibinfo {author}
  {\bibfnamefont {D.}~\bibnamefont {Sels}},\ }\bibfield  {title} {\enquote
  {\bibinfo {title} {Glassy phase of optimal quantum control},}\ }\href@noop {}
  {\bibfield  {journal} {\bibinfo  {journal} {Physical review letters}\
  }\textbf {\bibinfo {volume} {122}},\ \bibinfo {pages} {020601} (\bibinfo
  {year} {2019})}\BibitemShut {NoStop}%
\bibitem [{\citenamefont {Malmi-Kakkada}\ \emph {et~al.}(2019)\citenamefont
  {Malmi-Kakkada}, \citenamefont {Li}, \citenamefont {Sinha},\ and\
  \citenamefont {Thirumalai}}]{malmi2019dual}%
  \BibitemOpen
  \bibfield  {author} {\bibinfo {author} {\bibfnamefont {A.}~\bibnamefont
  {Malmi-Kakkada}}, \bibinfo {author} {\bibfnamefont {X.}~\bibnamefont {Li}},
  \bibinfo {author} {\bibfnamefont {S.}~\bibnamefont {Sinha}}, \ and\ \bibinfo
  {author} {\bibfnamefont {D.}~\bibnamefont {Thirumalai}},\ }\bibfield  {title}
  {\enquote {\bibinfo {title} {Dual role of cell-cell adhesion in tumor
  suppression and proliferation},}\ }\href@noop {} {\bibfield  {journal}
  {\bibinfo  {journal} {arXiv preprint arXiv:1906.11292}\ } (\bibinfo {year}
  {2019})}\BibitemShut {NoStop}%
\bibitem [{\citenamefont {Samanta}, \citenamefont {Sinha},\ and\ \citenamefont
  {Thirumalai}(2020)}]{samanta2020far}%
  \BibitemOpen
  \bibfield  {author} {\bibinfo {author} {\bibfnamefont {H.~S.}\ \bibnamefont
  {Samanta}}, \bibinfo {author} {\bibfnamefont {S.}~\bibnamefont {Sinha}}, \
  and\ \bibinfo {author} {\bibfnamefont {D.}~\bibnamefont {Thirumalai}},\
  }\bibfield  {title} {\enquote {\bibinfo {title} {Far from equilibrium
  dynamics of tracer particles embedded in a growing multicellular spheroid},}\
  }\href@noop {} {\bibfield  {journal} {\bibinfo  {journal} {arXiv preprint
  arXiv:2003.12941}\ } (\bibinfo {year} {2020})}\BibitemShut {NoStop}%
\bibitem [{\citenamefont {Wattenberg}, \citenamefont {Vi{\'e}gas},\ and\
  \citenamefont {Johnson}(2016)}]{wattenberg2016use}%
  \BibitemOpen
  \bibfield  {author} {\bibinfo {author} {\bibfnamefont {M.}~\bibnamefont
  {Wattenberg}}, \bibinfo {author} {\bibfnamefont {F.}~\bibnamefont
  {Vi{\'e}gas}}, \ and\ \bibinfo {author} {\bibfnamefont {I.}~\bibnamefont
  {Johnson}},\ }\bibfield  {title} {\enquote {\bibinfo {title} {How to use
  t-sne effectively},}\ }\href@noop {} {\bibfield  {journal} {\bibinfo
  {journal} {Distill}\ }\textbf {\bibinfo {volume} {1}},\ \bibinfo {pages} {e2}
  (\bibinfo {year} {2016})}\BibitemShut {NoStop}%
\bibitem [{\citenamefont {Almendro}, \citenamefont {Marusyk},\ and\
  \citenamefont {Polyak}(2013)}]{Alemendro13ARPathol}%
  \BibitemOpen
  \bibfield  {author} {\bibinfo {author} {\bibfnamefont {V.}~\bibnamefont
  {Almendro}}, \bibinfo {author} {\bibfnamefont {A.}~\bibnamefont {Marusyk}}, \
  and\ \bibinfo {author} {\bibfnamefont {K.}~\bibnamefont {Polyak}},\
  }\bibfield  {title} {\enquote {\bibinfo {title} {{Cellular Heterogeneity and
  Molecular Evolution in Cancer}},}\ }\href@noop {} {\bibfield  {journal}
  {\bibinfo  {journal} {Ann. Rev. Pathol. Med. Dis.}\ }\textbf {\bibinfo
  {volume} {8}},\ \bibinfo {pages} {277--302} (\bibinfo {year}
  {2013})}\BibitemShut {NoStop}%
\bibitem [{\citenamefont {Li}\ and\ \citenamefont
  {Thirumalai}(2019)}]{li2019share}%
  \BibitemOpen
  \bibfield  {author} {\bibinfo {author} {\bibfnamefont {X.}~\bibnamefont
  {Li}}\ and\ \bibinfo {author} {\bibfnamefont {D.}~\bibnamefont
  {Thirumalai}},\ }\bibfield  {title} {\enquote {\bibinfo {title} {Share, but
  unequally: a plausible mechanism for emergence and maintenance of intratumour
  heterogeneity},}\ }\href@noop {} {\bibfield  {journal} {\bibinfo  {journal}
  {Journal of the Royal Society Interface}\ }\textbf {\bibinfo {volume} {16}},\
  \bibinfo {pages} {20180820} (\bibinfo {year} {2019})}\BibitemShut {NoStop}%
\bibitem [{\citenamefont {Dolega}\ \emph {et~al.}(2017)\citenamefont {Dolega},
  \citenamefont {Delarue}, \citenamefont {Ingremeau}, \citenamefont {Prost},
  \citenamefont {Delon},\ and\ \citenamefont {Cappello}}]{Dolega17NC}%
  \BibitemOpen
  \bibfield  {author} {\bibinfo {author} {\bibfnamefont {M.~E.}\ \bibnamefont
  {Dolega}}, \bibinfo {author} {\bibfnamefont {M.}~\bibnamefont {Delarue}},
  \bibinfo {author} {\bibfnamefont {F.}~\bibnamefont {Ingremeau}}, \bibinfo
  {author} {\bibfnamefont {J.}~\bibnamefont {Prost}}, \bibinfo {author}
  {\bibfnamefont {A.}~\bibnamefont {Delon}}, \ and\ \bibinfo {author}
  {\bibfnamefont {G.}~\bibnamefont {Cappello}},\ }\bibfield  {title} {\enquote
  {\bibinfo {title} {Cell-like pressure sensors reveal increase of mechanical
  stress towards the core of multicellular spheroids under compression},}\
  }\href@noop {} {\bibfield  {journal} {\bibinfo  {journal} {Nature
  Communications}\ }\textbf {\bibinfo {volume} {8}},\ \bibinfo {pages} {14056}
  (\bibinfo {year} {2017})}\BibitemShut {NoStop}%
\bibitem [{\citenamefont {Alessandri}\ \emph {et~al.}(2013)\citenamefont
  {Alessandri}, \citenamefont {Sarangi}, \citenamefont {Gurchenkov},
  \citenamefont {Sinha}, \citenamefont {Kie{\ss}ling}, \citenamefont {Fetler},
  \citenamefont {Rico}, \citenamefont {Scheuring}, \citenamefont {Lamaze},
  \citenamefont {Simon} \emph {et~al.}}]{alessandri2013cellular}%
  \BibitemOpen
  \bibfield  {author} {\bibinfo {author} {\bibfnamefont {K.}~\bibnamefont
  {Alessandri}}, \bibinfo {author} {\bibfnamefont {B.~R.}\ \bibnamefont
  {Sarangi}}, \bibinfo {author} {\bibfnamefont {V.~V.}\ \bibnamefont
  {Gurchenkov}}, \bibinfo {author} {\bibfnamefont {B.}~\bibnamefont {Sinha}},
  \bibinfo {author} {\bibfnamefont {T.~R.}\ \bibnamefont {Kie{\ss}ling}},
  \bibinfo {author} {\bibfnamefont {L.}~\bibnamefont {Fetler}}, \bibinfo
  {author} {\bibfnamefont {F.}~\bibnamefont {Rico}}, \bibinfo {author}
  {\bibfnamefont {S.}~\bibnamefont {Scheuring}}, \bibinfo {author}
  {\bibfnamefont {C.}~\bibnamefont {Lamaze}}, \bibinfo {author} {\bibfnamefont
  {A.}~\bibnamefont {Simon}},  \emph {et~al.},\ }\bibfield  {title} {\enquote
  {\bibinfo {title} {Cellular capsules as a tool for multicellular spheroid
  production and for investigating the mechanics of tumor progression in
  vitro},}\ }\href@noop {} {\bibfield  {journal} {\bibinfo  {journal}
  {Proceedings of the National Academy of Sciences}\ }\textbf {\bibinfo
  {volume} {110}},\ \bibinfo {pages} {14843--14848} (\bibinfo {year}
  {2013})}\BibitemShut {NoStop}%
\bibitem [{\citenamefont {Doostmohammadi}\ \emph {et~al.}(2015)\citenamefont
  {Doostmohammadi}, \citenamefont {Thampi}, \citenamefont {Saw}, \citenamefont
  {Lim}, \citenamefont {Ladoux},\ and\ \citenamefont
  {Yeomans}}]{doostmohammadi2015celebrating}%
  \BibitemOpen
  \bibfield  {author} {\bibinfo {author} {\bibfnamefont {A.}~\bibnamefont
  {Doostmohammadi}}, \bibinfo {author} {\bibfnamefont {S.~P.}\ \bibnamefont
  {Thampi}}, \bibinfo {author} {\bibfnamefont {T.~B.}\ \bibnamefont {Saw}},
  \bibinfo {author} {\bibfnamefont {C.~T.}\ \bibnamefont {Lim}}, \bibinfo
  {author} {\bibfnamefont {B.}~\bibnamefont {Ladoux}}, \ and\ \bibinfo {author}
  {\bibfnamefont {J.~M.}\ \bibnamefont {Yeomans}},\ }\bibfield  {title}
  {\enquote {\bibinfo {title} {Celebrating soft matter's 10th anniversary: cell
  division: a source of active stress in cellular monolayers},}\ }\href@noop {}
  {\bibfield  {journal} {\bibinfo  {journal} {Soft Matter}\ }\textbf {\bibinfo
  {volume} {11}},\ \bibinfo {pages} {7328--7336} (\bibinfo {year}
  {2015})}\BibitemShut {NoStop}%
\bibitem [{\citenamefont {Puliafito}\ \emph {et~al.}(2012)\citenamefont
  {Puliafito}, \citenamefont {Hufnagel}, \citenamefont {Neveu}, \citenamefont
  {Streichan}, \citenamefont {Sigal}, \citenamefont {Fygenson},\ and\
  \citenamefont {Shraiman}}]{puliafito2012collective}%
  \BibitemOpen
  \bibfield  {author} {\bibinfo {author} {\bibfnamefont {A.}~\bibnamefont
  {Puliafito}}, \bibinfo {author} {\bibfnamefont {L.}~\bibnamefont {Hufnagel}},
  \bibinfo {author} {\bibfnamefont {P.}~\bibnamefont {Neveu}}, \bibinfo
  {author} {\bibfnamefont {S.}~\bibnamefont {Streichan}}, \bibinfo {author}
  {\bibfnamefont {A.}~\bibnamefont {Sigal}}, \bibinfo {author} {\bibfnamefont
  {D.~K.}\ \bibnamefont {Fygenson}}, \ and\ \bibinfo {author} {\bibfnamefont
  {B.~I.}\ \bibnamefont {Shraiman}},\ }\bibfield  {title} {\enquote {\bibinfo
  {title} {Collective and single cell behavior in epithelial contact
  inhibition},}\ }\href@noop {} {\bibfield  {journal} {\bibinfo  {journal}
  {Proceedings of the National Academy of Sciences}\ }\textbf {\bibinfo
  {volume} {109}},\ \bibinfo {pages} {739--744} (\bibinfo {year}
  {2012})}\BibitemShut {NoStop}%
\bibitem [{\citenamefont {Merkel}\ and\ \citenamefont
  {Manning}(2018)}]{merkel2018geometrically}%
  \BibitemOpen
  \bibfield  {author} {\bibinfo {author} {\bibfnamefont {M.}~\bibnamefont
  {Merkel}}\ and\ \bibinfo {author} {\bibfnamefont {M.~L.}\ \bibnamefont
  {Manning}},\ }\bibfield  {title} {\enquote {\bibinfo {title} {A geometrically
  controlled rigidity transition in a model for confluent 3d tissues},}\
  }\href@noop {} {\bibfield  {journal} {\bibinfo  {journal} {New Journal of
  Physics}\ }\textbf {\bibinfo {volume} {20}},\ \bibinfo {pages} {022002}
  (\bibinfo {year} {2018})}\BibitemShut {NoStop}%
\bibitem [{\citenamefont {Bi}\ \emph {et~al.}(2016)\citenamefont {Bi},
  \citenamefont {Yang}, \citenamefont {Marchetti},\ and\ \citenamefont
  {Manning}}]{Bi16PRX}%
  \BibitemOpen
  \bibfield  {author} {\bibinfo {author} {\bibfnamefont {D.}~\bibnamefont
  {Bi}}, \bibinfo {author} {\bibfnamefont {X.}~\bibnamefont {Yang}}, \bibinfo
  {author} {\bibfnamefont {M.~C.}\ \bibnamefont {Marchetti}}, \ and\ \bibinfo
  {author} {\bibfnamefont {M.~L.}\ \bibnamefont {Manning}},\ }\bibfield
  {title} {\enquote {\bibinfo {title} {Motility-driven glass and jamming
  transitions in biological tissues},}\ }\href@noop {} {\bibfield  {journal}
  {\bibinfo  {journal} {Physical Review X}\ }\textbf {\bibinfo {volume} {6}},\
  \bibinfo {pages} {021011} (\bibinfo {year} {2016})}\BibitemShut {NoStop}%
\bibitem [{\citenamefont {Altschuler}\ and\ \citenamefont
  {Wu}(2010)}]{altschuler2010cellular}%
  \BibitemOpen
  \bibfield  {author} {\bibinfo {author} {\bibfnamefont {S.~J.}\ \bibnamefont
  {Altschuler}}\ and\ \bibinfo {author} {\bibfnamefont {L.~F.}\ \bibnamefont
  {Wu}},\ }\bibfield  {title} {\enquote {\bibinfo {title} {Cellular
  heterogeneity: do differences make a difference?}}\ }\href@noop {} {\bibfield
   {journal} {\bibinfo  {journal} {Cell}\ }\textbf {\bibinfo {volume} {141}},\
  \bibinfo {pages} {559--563} (\bibinfo {year} {2010})}\BibitemShut {NoStop}%
\end{thebibliography}%

\end{document}